\documentclass[11pt]{article}
\usepackage{UF_FRED_paper_style}

\hypersetup{
    colorlinks = true,
    citecolor = blue
}
\usepackage{lipsum}  %% Package to create dummy text (comment or erase before start)
\usepackage{setspace}
\usepackage{amsmath}
\usepackage{amssymb}
\usepackage{bm}
\usepackage{booktabs}
\usepackage{subcaption}
\title{The Full Bayesian Significance Test and the e-value -- Foundations, theory and application in the cognitive sciences
}

\author{Riko Kelter\thanks{Correspondence concerning this article should be addressed to Riko Kelter, Department of Mathematics, University of Siegen, Walter-Flex-Street 3, 57072 Siegen, Germany.
    E-mail: riko.kelter$@$uni-siegen.de.
    Draft version 1.0, 2/6/20. This paper has not been peer reviewed. Please do not copy or cite without author's permission.}\\% Name author
	Department of Mathematics\\
	University of Siegen\\
    \and Julio Michael Stern\\
    Institute of Mathematics and Statistics\\
    University of S$\tilde{\text{a}}$o Paulo}

\date{\today}

\begin{document}
% %%%%%%%%%%%%%%%%%%%%%%%%%%%%%%%%%%%%%%%%%%%%%%%%%%%%%%%%%%
% %%%%%%%%%%%%%%%%%%%%%%%%%%%%%%%%%%%%%%%%%%%%%%%%%%%%%%%%%%
% ABSTRACT
% %%%%%%%%%%%%%%%%%%%%%%%%%%%%%%%%%%%%%%%%%%%%%%%%%%%%%%%%%%
% %%%%%%%%%%%%%%%%%%%%%%%%%%%%%%%%%%%%%%%%%%%%%%%%%%%%%%%%%%
{\setstretch{.8}
\maketitle
% %%%%%%%%%%%%%%%%%%
\begin{abstract}
% CONTENT OF ABS HERE--------------------------------------

Hypothesis testing is a central statistical method in psychological research and the cognitive sciences. While the problems of null hypothesis significance testing (NHST) have been debated widely, few attractive alternatives exist. In this paper, we provide a tutorial on the Full Bayesian Significance Test (FBST) and the $e$-value, which is a fully Bayesian alternative to traditional significance tests which rely on $p$-values. The FBST is an advanced methodological procedure which can be applied to several areas. In this tutorial, we showcase with two examples of widely used statistical methods in psychological research how the FBST can be used in practice, provide researchers with explicit guidelines on how to conduct it and make available $R$-code to reproduce all results. The FBST is an innovative method which has clearly demonstrated to perform better than frequentist significance testing. However, to our best knowledge, it has not been used so far in the psychological sciences and should be of wide interest to a broad range of researchers in psychology and the cognitive sciences.

% END CONTENT ABS------------------------------------------
\noindent
\textit{\textbf{Keywords: }%
Full Bayesian Significance Test; $e$-value; Bayesian hypothesis testing; significance testing} \\ %% <-- Keywords HERE!
\noindent
%\textit{\textbf{JEL Classification: }%
%Q12; C22; D81.} %% <-- JEL code HERE!

\end{abstract}
}

% %%%%%%%%%%%%%%%%%%%%%%%%%%%%%%%%%%%%%%%%%%%%%%%%%%%%%%%%%%
% %%%%%%%%%%%%%%%%%%%%%%%%%%%%%%%%%%%%%%%%%%%%%%%%%%%%%%%%%%
% BODY OF THE DOCUMENT
% %%%%%%%%%%%%%%%%%%%%%%%%%%%%%%%%%%%%%%%%%%%%%%%%%%%%%%%%%%
% %%%%%%%%%%%%%%%%%%%%%%%%%%%%%%%%%%%%%%%%%%%%%%%%%%%%%%%%%%

% --------------------
Hypothesis testing is a central statistical method in psychological research. While the problems of null hypothesis significance testing (NHST) have been debated widely, few attractive alternatives exist for practitioners. In this paper, we provide a tutorial on the Full Bayesian Significance Test (FBST) and the $e$-value, which is a fully Bayesian alternative to traditional significance tests. We show that the FBST is an advanced methodological procedure which can be applied to several areas of psychological research. Two examples of widely used statistical methods in psychological research highlight how the FBST can be used in practice, and we provide researchers with explicit guidelines how to conduct it by providing $R$-code to reproduce all results and analyses. The FBST is an innovative method which has clearly demonstrated to perform better than frequentist significance testing. However, to our best knowledge, it has not been used so far in the psychological sciences and should be of wide interest to a broad range of researchers in psychology.

The last century has brought the advent of multiple proposals on how to test a research hypothesis statistically \citep{Howie2002}. Well-known examples include the theory of significance testing employing $p$-values, formally introduced by British statistician Ronald Fisher \citep{Fisher1925} and the theory of uniformly most powerful tests of \cite{Neyman1933}. While both theories can differ substantially in the application and obtained results \citep{Cox1958}, they are unified by being located under the umbrella of the frequentist statistical philosophy \citep{Mayo2018}. The more recent replication crisis in psychology and its relationship to the frequentist paradigm, in particular, has been discussed widely in the last decade \citep{Pashler2012a,Colquhoun2014,OpenScienceFoundation}. While the problems of null hypothesis significance testing (NHST) and $p$-values have been analysed and detailed in various articles \citep{Colquhoun2016,Colquhoun2017,Greenland2016,Greenland2019}, the experienced reproducibility issues are far from being solved \citep{Ioannidis2019,Matthews2017}. 

In general, among the proposed solutions to the observed problems with NHST and $p$-values is a trend for the increased use of Bayesian data analysis \citep{wasserstein2016,Wasserstein2019}. Narrowing the scope to psychological research, there is an increasing trend of proposals which recommend a shift towards Bayesian statistics, in particular towards Bayesian hypothesis testing \citep{Rouder2014,Morey2016,Wagenmakers2016}. Often, these proposals centre on the Bayes factor as a replacement for traditional $p$-values \citep{Wagenmakers2010,Hoijtink2019}, and emphasize the benefits of Bayesian interval estimates over traditional confidence intervals \citep{Kruschke2018b,Morey2016c,Wagenmakers2020}. However, there are also discussions about the benefits of applying Bayesian data analysis in meta-analysis or clinical trials \citep{Kruschke2018}.

Considering a wider timeframe for a moment reveals that Bayesian mathematical psychology has become more popular in the last decades in general. \cite{VanDeSchoot2017} conducted an extensive systematic review which included $n=1579$ Bayesian psychologic articles published between $1990$ and $2015$, and concluded that Bayesian statistics ``is used in a variety of contexts across subfields of psychology and related disciplines.'' \cite[p.~1]{VanDeSchoot2017}. They underlined that
\begin{quote}
    ``There are many different reasons why one might choose to use Bayes (e.g., the use of priors, estimating otherwise intractable models, modeling uncertainty, etc.). We found in this review that the use of Bayes has increased and broadened in the sense that this methodology can be used in a flexible manner to tackle many different forms of questions.''\newline
    \cite[p.~1]{VanDeSchoot2017}
\end{quote}
However, while there is a trend which favours Bayesian data analysis over frequentist solutions in mathematical psychology \citep{Dienes2018}, there are also critical voices. \cite{Tendeiro2019} recently reviewed some issues about the practice of Bayesian hypothesis testing via Bayes factors which are often advocated in the literature \cite{Morey2016}. Among the problems discussed are (1) the sensitivity of Bayes factors to within-model priors \citep{Kamary2014,Robert2016,Kelter2020BMCJasp}, (2) the requirement of mathematically advanced numerical methods for the computation of Bayes factors like the \textit{Savage-Dickey density ratio method} \citep{Dickey1970,Verdinelli1995,Wagenmakers2010} or \textit{bridge sampling} \citep{Gronau2017,Gronau2019}, and (3) the fact that the thresholds for interpreting Bayes factors are similarly arbitrary as the significance levels used on frequentist hypothesis tests \citep{Tendeiro2019}. For more details on these issues see \cite{Tendeiro2019}. Other authors even argue in favour of NHST and $p$-values and only criticise that practitioners use and interpret them inappropriately \citep{Greenland2019}. Recent results also have shown that there are various other Bayesian indices as the Bayes factor for significance and the size of an effect, some of which have appealing theoretical and practical properties \citep{Makowski2019,Kelter2020BayesianPosteriorIndices}. This situation shows that it is useful to widen the scope in the discussion about statistical significance when it comes to Bayesian hypothesis testing.

In summary, the existing literature indicates that there is no trivial solution to the methodological status quo, which in psychology is still based on NHST and $p$-values \citep{Matthews2017}, and it is not the goal of this paper to join the discussion ``Bayes factors vs. $p$-values'' for hypothesis testing in psychological research. Instead, the goal of this paper is to draw attention to a statistical method which has despite its various appealing properties -- to the author's best knowledge -- not been applied in psychology so far. While it offers an appealing alternative to contemporary statistical approaches to hypothesis testing, the roots of the procedure date back more than two decades and it has been applied successfully in a wide range of scientific areas to the present date.

\section{The Full Bayesian Significance Test}
\subsection{The philosophy behind the FBST}
This section outlines the theory of the \textit{Full Bayesian Significance Test (FBST)} and the $e$-value, which enjoys desirable properties and is easy to apply in practice. The Full Bayesian Significance Test was developed more than two decades ago by \cite{Pereira1999} as a fully Bayesian alternative to traditional frequentist null hypothesis significance tests. It was designed to test a \texttt{sharp} (or precise) point null hypothesis $H_0$ against its alternative $H_1$.

According to \cite{Cox1977} and \cite{Kempthorne1976}, a significance test is defined as a method which measures the consistency of data with a null hypothesis $H_0$. Frequentist hypothesis tests use $p$-values which are based on the idea of ordering the \textit{sample space} according to increasing inconsistency with the hypothesis. In contrast, the e-value used in the FBST is based on the idea of ordering the \textit{parameter space} according to increasing inconsistency with observed data \citep{Pereira2008}. Traditional frequentist significance testing employs the $p$-value to reject the null hypothesis $H_0$:
\begin{align*}
    p=Pr(x\in C|\theta_0)
\end{align*}
Here, often $C:=\{x\in \mathcal{X}|T_{\theta_0}(x)\geq t_{\text{obs}}\}$ is the set of sample space values $x\in \mathcal{X}$ for which a test statistic $T_{\theta_0}$ under assumption of the null hypothesis value $\theta_0$ is at least as large as the test statistic value $t_{\text{obs}}$ calculated from the observed data. The set $C$ is, in general, interpreted as the sample space values $x\in \mathcal{X}$ which are at least as \textit{inconsistent} with the null hypothesis value $\theta_0$ as the observed data, and the $p$-value quantifies the evidence against $H_0$ by calculating the probability over this set \citep{Casella2002a,Pereira2008,Held2014}. 

\cite[p.~80]{Pereira2008} argued that a Bayesian should look at the so-called \textit{tangential set} $T$ of parameter points which are \textit{more consistent} with the data $x$ than $\theta_0$, that is at $ev = 1-\overline{ev}$, where
\begin{align*}
    \overline{ev} = Pr(\theta \in T|x)
\end{align*}
Here, $ev$ is interpreted as the evidence in favour of $H_0$, while $\overline{ev}$ is interpreted as the evidence against $H_0$, which is the probability of all parameter values $\theta$ in the parameter space $\Theta$ that are \textit{more consistent} with the data $x$ than the null value $\theta_0$. The philosophy of the FBST is, in summary, based on constructing a duality between sampling theories and Bayesian theory. More precisely, the philosophy of the FBST is based on the duality between frequentist significance measures based on an incompatibility order defined in the sample space, and the Bayesian e-value based on an incompatibility order defined in the parameter space. Notice that a frequentist likelihood ratio test compares the supremum of the likelihood restricted to the null set with the supremum of the likelihood under the alternative to measure the inconsistency of the data with the null hypothesis. In the FBST, the tangential set is based on the posterior distribution, allowing a Bayesian perspective. Also, the tangential set is, as the name says, a set of values instead of a supremum under a hypothesis, which produces a less subjective statement of evidence \citep{Berger1987}. Before the next section outlines the mathematical theory behind the FBST in more detail, notice that the consequences of this philosophical basis are substantial: The quantity $\overline{ev}$ is not a mere Bayesian counterpart to the frequentist $p$-value, but a genuine Bayesian procedure in the sense that it follows the likelihood principle \citep{Birnbaum1962,Basu1975,Berger1988a}. As a consequence, among the advantages of using the FBST is:
\begin{itemize}
    \item{Researchers are allowed to make use of optional stopping. This implies that it is permitted to stop recruiting participants or abort an experiment and report the results when only a fraction of the observed data shows overwhelming evidence. Notice that this behaviour results in severe problems when NHST and $p$-values are used. For a review for psychologists see \cite{Edwards1963}, \cite{Kruschke2015} and \cite{Kruschke2018}.}
    \item{The interpretation of censored data (which is often observed in longitudinal studies or clinical trials) is conceptually simplified \citep{Berger1988a}. As a consequence of the likelihood principle, the likelihood contribution of a single observation in an experiment where no censoring is possible is equal to the likelihood contribution of a single observation in an experiment where censoring is possible but did not happen to the observation, see \cite[Chapter 4]{Berger1988a}.}
    \item{The results obtained do not depend on the researcher's intentions \citep{Kruschke2018a}.}
\end{itemize}
These aspects are already appealing to practitioners. Besides, the FBST can be formally derived as a Bayes rule, which means that it can be derived by minimising an appropriate loss function \citep{Madruga2001,Madruga2003,DaSilva2015}. Also, the FBST has logical properties which are not met by both frequentist $p$-values and Bayes factors. We only mention two of these here, but for more details see \cite{Stern2003}. 

First, most (frequentist or Bayesian) hypothesis testing approaches try to express the support for a sharp null hypothesis $H_0:\theta=\theta_0$ via the probability of the null set. For a sharp hypothesis $H_0:\theta=\theta_0$, the null set is simply the point $\{\theta_0\}$ which has Lebesgue measure zero \citep{Bauer2001}. As noted by \cite[p.~5]{Stern2003}, to prevent measure-theoretic problems when assigning a prior probability to a set of measure zero, various statistical tests reparameterize the hypothesis in a specific way and then use a probability measure on the submanifold derived by reparameterizing the hypothesis in this specific way. As a consequence, the probability measure used for quantifying the evidence does not operate in the original parameter space. In contrast, the probability measure used with the FBST does \cite[Section 3]{Stern2020}.

Second, various approaches to sharp hypothesis testing use the nuisance parameter elimination paradigm. For example, the Bayes factor $BF_{01}$ in favor of $H_0$ is the ratio of the marginal likelihoods $p(x|H_0)$ and $p(x|H_1)$:
\begin{align*}\label{eq:bf}
    \underbrace{\frac{\mathbb{P}(H_0|x)}{\mathbb{P}(H_1|x)}}_{\text{Posterior odds}}    =\underbrace{\frac{p(x|H_0)}{p(x|H_1)}}_{BF_{01}(x)}\cdot \underbrace{\frac{\mathbb{P}(H_0)}{\mathbb{P}(H_1)}}_{\text{Prior odds}}
\end{align*}
However, the marginal likelihoods can be difficult to obtain in the presence of nuisance parameters which are not of interest for the problem at hand. To obtain the marginal likelihood $p(x|H_1)$, these have to be integrated out
\begin{align*}
    p(x|H_1)=\int p(x|\bm{\theta},H_1)p(\bm{\theta}|H_1)d\bm{\theta}
\end{align*}
where the parameter $\bm{\theta}$ is possible vector-valued. This quickly becomes cumbersome in high-dimensional models and troubles inference further \citep{Rubin1984,Held2014}. In contrast, the FBST operates in the original parameter space and does not need to eliminate nuisance parameters to be conducted \citep{Pereira2008}. This latter property makes application of the FBST straightforward as will be shown in the examples later.

\subsection{The mathematical theory behind the FBST}
The previous section gave an overview of the foundational aspects of the FBST. The adherence to the likelihood principle is a substantial benefit of the FBST compared to NHST which relies on $p$-values. Also, there are some conceptual simplifications over other indices like Bayes factors or $p$-values. This section describes the mathematical theory behind the FBST in more detail and reveals that these simplifications also have their price. Importantly, this section shows that in contrast to the Bayes factor, the FBST can \textit{not} confirm a research hypothesis. Via the FBST, one can only state evidence \textit{against} a sharp null hypothesis $H_0$, which shows the similarity to the frequentist $p$-value.

Nevertheless, the FBST can be generalized into an extended framework which allows for hypothesis confirmation \citep{Esteves2019}. This wider framework was constructed because precise hypotheses cannot be accepted by logically consistent tests. \cite{Esteves2019} showed that this dilemma can be overcome by the use of pragmatic versions of precise hypotheses, which allows a level of imprecision in the hypothesis that is small relative to other experimental conditions. The introduction of pragmatic hypotheses in turn allows the evolution of scientific theories based on statistical hypothesis testing and the FBST can be generalized into this theory.\footnote{Notice the strong analogy to often proposed approaches of equivalence testing in contemporary mathematical psychology, see \cite{Lakens2017,Lakens2018}, \cite{Kruschke2018b,Kruschke2018a} or \cite{Liao2020}.}

The FBST can be used with any standard parametric statistical model, where $\theta \in \Theta \subseteq \mathbb{R}^p$ is a (vector-valued) parameter of interest, $p(x|\theta)$ is the model likelihood and $p(\theta)$ is the prior distribution for the parameter $\theta$ of interest. A sharp (or expressed equivalently, precise) hypothesis $H_0$ makes a statement about the parameter $\theta$: Specifically, the null hypothesis $H_0$ states that $\theta$ lies in the so-called \textit{null set} $\Theta_{H_0}$. For simple point null hypotheses like $H_0:\theta=\theta_0$ often used in practice this null set is just the single parameter value $\theta_0$, so that the null set can be written as $\Theta_{H_0} = \{\theta_0 \}$. As detailed in the previous section, the approach of the FBST consists of stating the Bayesian evidence against $H_0$, the $e$-value. This value is the proposed Bayesian replacement of the traditional $p$-value. To construct the $e$-value, \cite{Pereira2008} used the posterior \textit{surprise function} $s(\theta)$ which is defined as follows:
\begin{align}
    s(\theta):=\frac{p(\theta|x)}{r(\theta)} 
\end{align}
The surprise function $s(\theta)$ is the ratio of the posterior distribution $p(\theta|x)$ and a suitable \textit{reference function} $r(\theta)$. The first thing to note is that two important special cases are given by a flat reference function $r(\theta)=1$ or any prior distribution $p(\theta)$ for the parameter $\theta$. When a flat reference function is selected the surprise function recovers the posterior distribution $p(\theta|x)$. When any prior distribution is used as the reference function, one can interpret parameter values $\theta$ with a surprise function value of one or larger, that is with $s(\theta)\geq 1$, as being corroborated by observing the data $x$. In contrast, parameter values $\theta$ with a surprise function $s(\theta)<1$ then indicate that they have not been corroborated by observing the data. The next step is to calculate the supremum $s^{*}$ of the surprise function $s(\theta)$ over the null set $\Theta_{H_0}$.
\begin{align*}
    s^{*}:=s(\theta^{*})=\sup\limits_{\theta \in \Theta_{H_0}}s(\theta)
\end{align*}
This supremum is subsequently used in combination with the tangential set, which has been introduced in the last section already. \cite{Pereira2008} defined the tangential set $\overline{T}(\nu)$ to the sharp null hypothesis $H_0$ as follows:
\begin{align}
    \overline{T}(\nu):=\Theta \setminus T(\nu)
\end{align}
In the above, $T(\nu)$ is given as
\begin{align}\label{eq:tangentialSet}
    T(\nu):=\{\theta \in \Theta|s(\theta)\leq \nu \}
\end{align}
Using the value $\nu=s^{*}$, the tangential set $\overline{T}(\nu)$ has precisely the interpretation discussed in the previous section: $T(s^{*})$ then includes all parameter values $\theta$ which are either smaller or equal to the supremum value $s^{*}$ of the surprise function $s(\theta)$ over the null set $\Theta_{H_0}$. As a consequence of equation (\ref{eq:tangentialSet}), the tangential set $\overline{T}(s^{*})$ includes all parameter values $\theta$ which are \textit{larger} than the supremum $s^{*}$ of the surprise function over the null set $\Theta_{H_0}$.

\begin{figure*}[!h]
\centering
\includegraphics[width=0.9\textwidth]{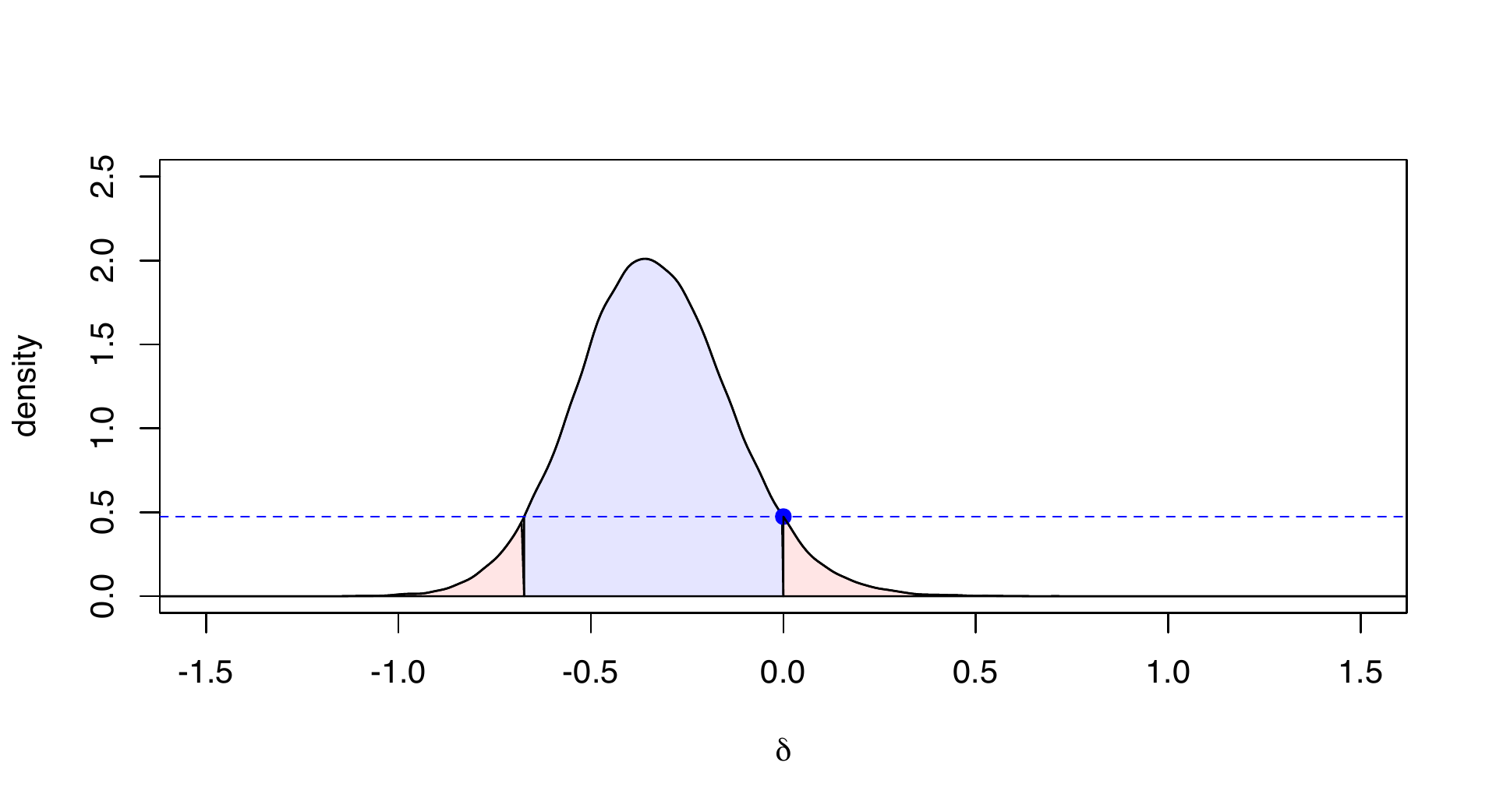}
\caption{Visualisation of the FBST and the $e$-value $\overline{\text{ev}}(H_0)$ against $H_0:\delta=0$ in a Bayesian two-sample t-test, where $\delta$ is the effect size. A flat reference function $r(\delta)=1$ is used, and the solid line is the resulting posterior distribution $p(\delta|x)$ after observing the data. The supremum over the null set is $s^{*}=0$, highlighted as the blue point. The blue shaded area is $\overline{W}(0)$, the integral over the tangential set $\overline{T}(0)$ of $H_0:\delta =0$, which is the $e$-value $\overline{\text{ev}}(H_0)$ against $H_0$; the red area is the integral $W(0)$ over $T(0)$, which is the $e$-value ev$(H_0)$ in favour of $H_0:\delta=0$}
\label{fig:example1}
\end{figure*}

\begin{figure*}[!h]
\centering
\includegraphics[width=0.9\textwidth]{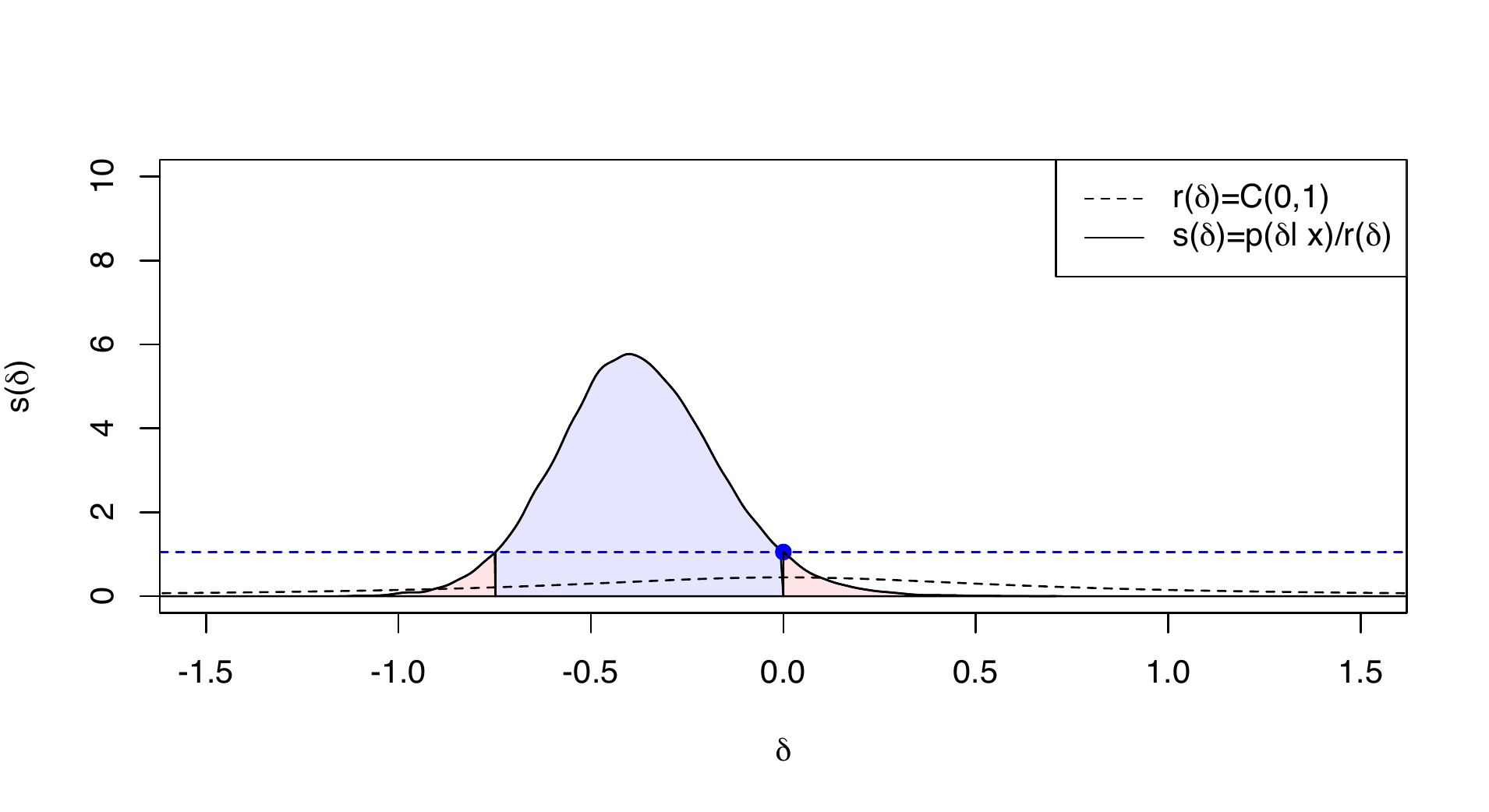}
\caption{Visualisation of the FBST and the $e$-value $\overline{\text{ev}}(H_0)$ against $H_0:\delta=0$ in a Bayesian two-sample t-test, where $\delta$ is the effect size. A Cauchy reference function $r(\delta)=C(0,1)$ is used, and the solid line is the resulting surprise function $p(\delta|x)/r(\delta)$ after observing the data $x$. The supremum over the null set is $s^{*}=0$, highlighted as the blue point. The blue shaded area is $\overline{W}(0)$, the integral over the tangential set $\overline{T}(0)$ against $H_0:\delta =0$, which is the $e$-value $\overline{\text{ev}}$ against $H_0$; the red area is the integral $W(0)$ over $T(0)$, which is the $e$-value ev$(H_0)$ in favour of $H_0:\delta=0$}
\label{fig:fbst}
\end{figure*}

Following the ideas presented in the previous section, the final step to obtain the $e$-value is to define the \textit{cumulative surprise function} $W(\nu)$
\begin{align}
    W(\nu):=\int_{T(\nu)}p(\theta|x)d\theta
\end{align}
The cumulative surprise function $W(\nu)$ is the integral of the posterior distribution $p(\theta|x)$ over all parameter values which have a surprise function value $s(\theta)\leq \nu$. Again, setting $\nu=s^{*}$, $W(s^{*})$ becomes the integral of the posterior distribution $p(\theta|x)$ over $T(s^{*})$, which is the integral of the posterior $p(\theta|x)$ over all parameter values which have a surprise function value $s(\theta)\leq s^{*}$. Finally, the Bayesian evidence against $H_0$, the $e$-value against $H_0$ is calculated as
\begin{align}
    \overline{\text{ev}}(H_0):=\overline{W}(s^*)
\end{align}
where $\overline{W}(\nu):=1-W(\nu)$. Figure \ref{fig:example1} shows the single parts which are used in the FBST and visualises the $e$-value $\overline{\text{ev}}(H_0)$. The solid line shows the posterior distribution of the effect size $\delta$ and is based on a Bayesian two-sample t-test \citep{Kelter2020JORSBayest}.
A flat reference function $r(\delta)=1$ was used, and the solid line is the resulting posterior distribution $p(\delta|x)$ after observing the data $x$. The supremum over the null set $\Theta_{H_0}=\{0\}$ is $s^{*}=s(0)$, highlighted as the blue point. The horizontal blue dashed line shows the boundary between $T(0)$ and $\overline{T}(0)$: Values with posterior density $p(\delta)>p(0)$ are in $\overline{T}(0)$, while values with $p(\delta)\leq p(0)$ are in $T(0)$. The blue shaded area is $\overline{W}(0)$, the integral over the tangential set $\overline{T}(0)$ against $H_0:\delta =0$, which is the $e$-value $\overline{\text{ev}}(H_0)$ against $H_0$; the red area is the integral $W(0)$ over $T(0)$, which is the $e$-value ev$(H_0)$ in favour of $H_0:\delta=0$. The resulting $e$-value against $H_0:\delta=0$ is given as  $\overline{\text{ev}}(H_0)=0.907$, which is the amount of probability mass shaded in blue in figure \ref{fig:example1}. Based on this value, there is considerable evidence against the null hypothesis. Now, instead of flat reference function $r(\delta)=1$ it is also possible to choose a proper prior distribution. Figure \ref{fig:fbst} visualises the same situation but now the reference function was selected as a medium Cauchy prior $C(0,1)$, which is often recommended in the setting of the Bayesian two-sample t-test \citep{Rouder2009}. In figure \ref{fig:fbst}, the reference function is shown as the dashed black line. The surprise function now does not become the posterior distribution. Instead, it is the ratio $s(\delta)=p(\delta|x)/r(\delta)$, where $r(\delta)$ is the Cauchy prior.

Notice that although the situation seems quite similar to figure \ref{fig:example1}, the scaling on the $y$-axis has changed. Also, the interpretation of the surprise function is now different. If one would assume a Cauchy prior $C(0,1)$ on the effect size $\delta$, parameters with a surprise function value $s(\delta)\geq 1$ can be interpreted as being corroborated by the data when such a prior distribution of the effect size is assumed. The resulting $e$-value against $H_0:\delta=0$ visualised as the shaded blue area is given as $\overline{\text{ev}}(H_0)=0.932$.

Formally, \cite{Pereira1999} defined the $e$-value ev$(H_0)$ in \textit{support} of $H_0$ as 
\begin{align}
    \text{ev}(H_0):=1-\overline{\text{ev}}(H_0)
\end{align}
Nevertheless, notice that the Bayesian evidence in support of $H_0$ (that is, the $e$-value $\text{ev}(H_0)$) can not be interpreted as evidence \textit{against} $H_1$. This is because $H_1$ is not even a sharp hypothesis, compare also with Definition 2.2 in \cite{Pereira2008}.

Importantly, it is not possible to utilise the $e$-value ev$(H_0)$ to \textit{confirm} the null hypothesis $H_0$. The reasons can be attributed to the fact that even when the posterior concentrates around the true value $\theta_0$ of the null hypothesis $H_0:\theta=\theta_0$, the posterior distribution's probability mass fluctuates around the true value according to the central limit theorem. For details see \cite{Kelter2020BayesianPosteriorIndices}. However, one can use ev$(H_0)$ to reject $H_0$ if ev$(H_0)$ is sufficiently small, and there are asymptotic arguments based on the distribution of ev$(H_0)$ \cite[Section 5]{Pereira2008}. \cite{Pereira2008} derived the distribution of the $e$-value as a Chi-square distribution
\begin{align}\label{eq:evPVal}
    \text{ev}(H_0) \sim \chi_k^2(||m-M||^2)    
\end{align}
where $M$ and $m$ are the posterior mode calculated over the entire parameter space $\Theta$ and the posterior maximum restricted to $\Theta_{H_0}$ (that is, $s^{*}$). They showed that the $p$-value associated with the Bayesian evidence in support of $H_0$, the $e$-value $\text{ev}(H_0)$, is the superior tail of the $\chi^2$ density with $k-h$ degrees of freedom, starting from $-2\lambda(m_0)$. Here, $k$ and $h$ are the dimensions of the parameter space $\Theta$ and the null set $\Theta_{H_0}$. $m_0$ is the observed value and $\lambda(t)=\ln l(t)$ where $l(t)=L(t)/L(M)$ is the relative likelihood. Now, the $p$-value associated with the Bayesian $e$-value $\text{ev}(H_0)$ is then given as
\begin{align}
    pv_{0}=1-\chi_{k-h}^2(-2\lambda(m_0))    
\end{align}
Notice that this latter $p$-value has a frequentist interpretation, while equivalently, the $p$-value based on equation (\ref{eq:evPVal}) can be expressed as
\begin{align}\label{eq:evPVal2}
    ev_0 = \chi_k^2(||m_0-M_0||^2)    
\end{align}
which is interpreted as a Bayesian significance value. As a consequence, when observing $m_0$ and $M_0$ which are the maximum restricted to $\Theta_{H_0}$ (that is, $s^{*}$) and the posterior mode, one only needs to calculate the euclidian distance $d_0=||m_0-M_0||^2$ and obtain the value of the $\chi_{k}^2$ distribution of this distance. Then, a usual threshold of the $\chi_k^2$ distribution can be used to reject the null hypothesis $H_0:\theta=\theta_0$ or not.

There is also the option to use the standardized $e$-value $\text{sev}(H_0)$, as defined in \cite[Section 2.2]{Borges2007} and in \cite[Section 3.3]{Pereira2020}, which is the quantity that constitutes the best analogue to a frequentist $p$-value. The standardized $e$-value is defined as: $\overline{\text{sev}}(H_0)=F_{k-h}(F^{-1}_{k}(\overline{\text{ev}}))$, where $F^{-1}_{k}$ is the generalised inverse of the cumulative distribution function of the $\chi_{k}^2$ distribution with $k$ degrees of freedom, and $F_{k-h}$ is the cumulative distribution function of the $\chi_{k-h}^2$ distribution. $\overline{\text{sev}}(H_0)$ can be interpreted as the probability of obtaining less evidence than $\overline{\text{ev}}(H_0)$ against the null hypothesis $H_0$. Using the relationship $\text{sev}(H_0)=1-\overline{\text{sev}}(H_0)$, $\text{sev}(H_0)$ can be interpreted as the probability of obtaining $\overline{\text{ev}}(H_0)$ or more evidence against $H_0$. Notice the strong analogy to the frequentist $p$-value, which is why this standardised $e$-value $\text{sev}(H_0)$ can be used as a replacement for frequentist $p$-values if desired. In the examples, we therefore report the raw Bayesian evidence against $H_0$, that is, $\overline{\text{ev}}(H_0)$, the normal $p$-value associated with $\overline{\text{ev}}(H_0)$, that is, $ev_0$, and also the standardized $e$-value $\text{sev}(H_0)$. Notice that when a $p$-value replacement is desired, the latter quantity $\text{sev}(H_0)$ is most suitable, for details see \cite{Pereira2020}.

In summary, the FBST and the $e$-value were invented to precisely mimic a frequentist significance test of a \textit{sharp} hypothesis. The $e$-value $\overline{\text{ev}}(H)$ can be interpreted as a direct replacement of the frequentist $p$-value and can \textit{only} be used to reject a null hypothesis $H_0$ of interest, either based on a continuous interpretation (which we follow here) or based on the asymptotic arguments outlined above. If the asymptotic arguments are used, the standardized $e$-value $\text{sev}(H_0)$ has the strongest similarity to a frequentist $p$-value, while $ev_0$ has a more Bayesian flavour. Notice, however, that the confirmation of a research hypothesis via ev$(H)$ is not possible via the FBST \citep{Kelter2020BayesianPosteriorIndices, Pereira2020}.

%\subsection{Applications of the FBST}
%Optional, vllt weglassen
\section{Examples and illustrations of the FBST for psychological methods}
This section provides two examples which show how to apply the FBST in practice. The first example is a Bayesian two-sample t-test \citep{Rouder2009}. The two-sample t-test is one of the most widely used statistical procedures carried out in psychological research \citep{Nuijten2016} and as a consequence allows readers to apply the FBST on their own t-tests if desired.

The second example is based on the linear regression model. Linear regression is also an important statistical method in psychology and the biomedical sciences \citep{VanErp2019,Faraway2016}, and here we focus on testing if a regression coefficient $\beta_j$ for a specific predictor is zero or not. That is, we test $H_0:\beta_j=0$ against $H_1:\beta_j \neq 0$ for a regression coefficient $\beta_j$.

While we prefer a continuous interpretation of the $e$-value in the same way we prefer a continuous interpretation of $p$-values we also provide the resulting $p$-values $ev_0$ associated with the $e$-value ev$(H_0)$ and the standardized $e$-values $\text{sev}(H_0)$ for the interested reader. Note however that the continuous quantification of evidence against $H_0$ based solely on $\overline{\text{ev}}(H_0)$ is not anymore arbitrary than the decision based on the $p$-value $ev_0$ associated with ev$(H_0)$ or the standardized $e$-value $\text{sev}(H_0)$: In the latter case, the decision threshold used for separating significant from non-significant $p$-values (like $ev_0 < .05$ or $\text{sev}(H_0)<.05$) is as arbitrary as using a threshold like $\overline{\text{ev}}(H_0)>0.95$ on $\overline{\text{ev}}(H_0)$ (or ev$(H_0)$) directly.

We encourage readers to reproduce all analyses via the provided replication script, which is available at the Open Science Foundation at \url{https://osf.io/8rg2k/}.

\subsection{The FBST in the setting of the Bayesian two-sample t-test}
In the first example we use data from \cite{Wagenmakers2015}, who replicated the study of \cite{Topolinski2012}.\footnote{The data is freely available in the built-in data library of the open-source statistical software JASP, freely available at \url{www.jasp-stats.org}.} In their paper called ``Turning the Hands of Time'', \cite{Topolinski2012} conducted a study in which participants were split into two groups and each group had to fill out a personality questionnaire measuring the openness to new experiences. The first group had to roll a kitchen roll counterwise while completing the questionnaire. In contrast, the second group had to roll the kitchen roll clockwise. The personality questionnaire's mean score was recorded for each participant and these are compared via the Bayesian two-sample t-test of \cite{Rouder2009}. Formally, we test the hypothesis $H_0:\delta=0$ against $H_1:\delta \neq 0$, which is equivalent to $H_0:\mu_1=\mu_2$ against $H_1:\mu_1 \neq \mu_2$ due to the definition of $\delta$, compare \cite{cohen_statistical_1988}.

\begin{figure}
	\begin{subfigure}[c]{0.50\textwidth}
		\includegraphics[width=1\textwidth]{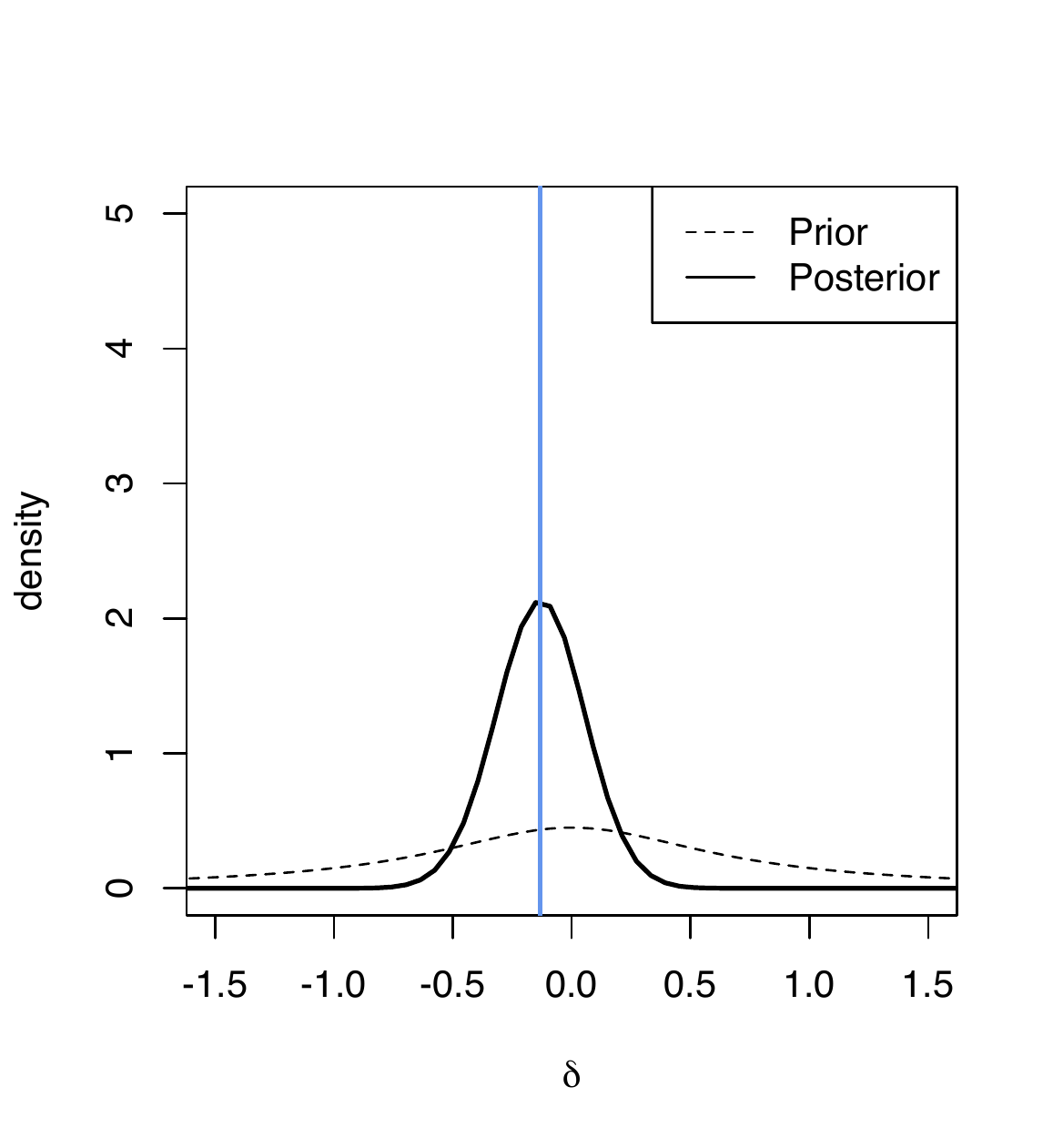}
		\subcaption{Prior-posterior plot of effect size $\delta$ in the Bayesian two-sample t-test for the kitchen rolls data set}
		\label{fig:example1posterior}
	\end{subfigure}
	\begin{subfigure}[c]{0.50\textwidth}
		\includegraphics[width=1\textwidth]{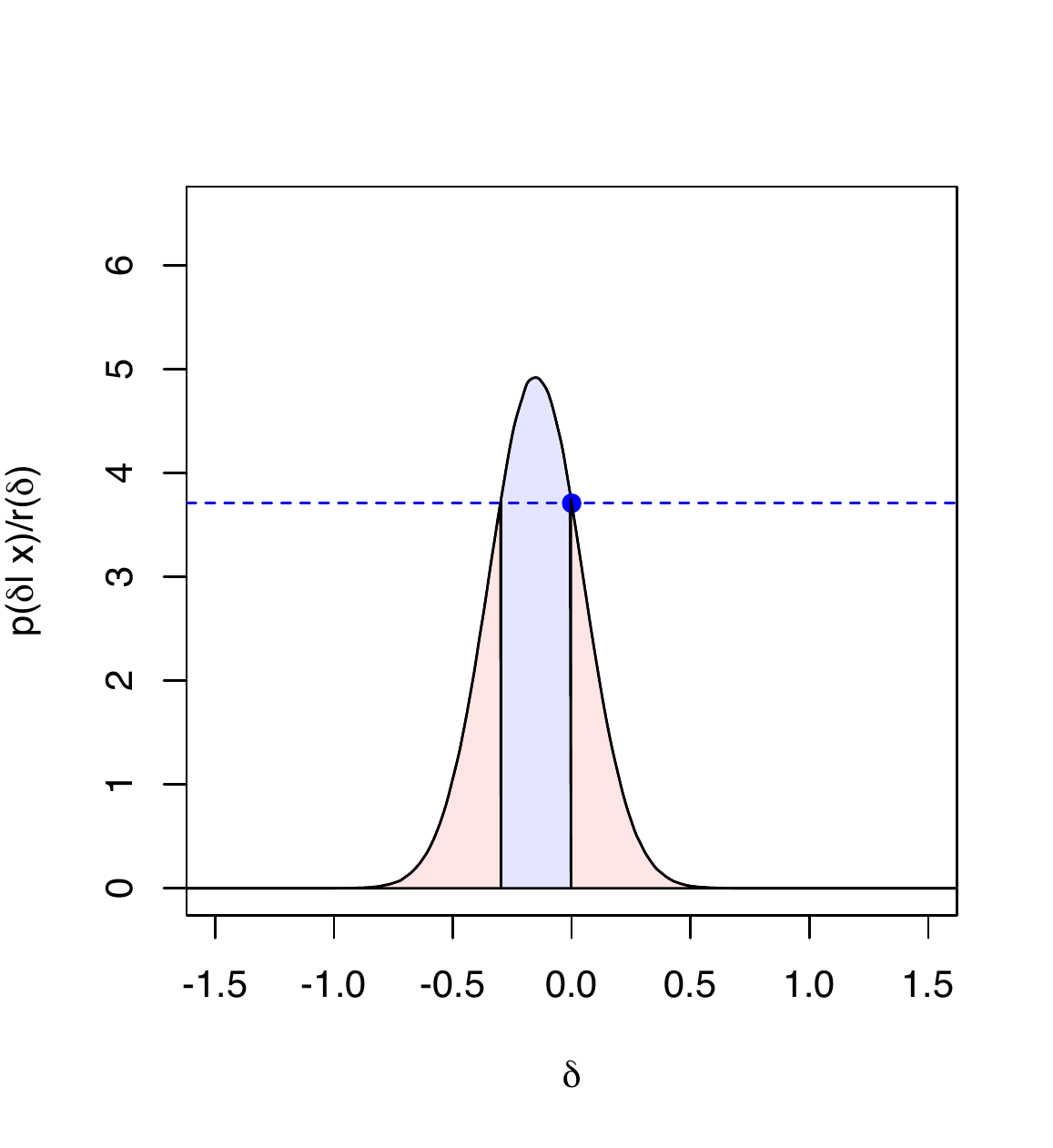}
		\subcaption{FBST for $H_0:\delta=0$ against $H_1:\delta \neq 0$ in the Bayesian two-sample t-test for the kitchen rolls data set using $r(\delta)=C(0,\sqrt{2}/2)$}
		\label{fig:example1fbst}
	\end{subfigure}
	\caption{Prior-posterior plot and FBST for the Bayesian two-sample t-test for the kitchen rolls dataset}
\end{figure}

Figure \ref{fig:example1posterior} shows the resulting prior-posterior plot of the effect size $\delta$ for the Bayesian two-sample t-test based on the observed data in both groups. The recommended medium Cauchy prior $C(0,\sqrt{2}/2)$ was assigned to the effect size \citep{Rouder2009}. The posterior distribution has moved from the prior distribution which is centred at zero towards negative effect sizes and the 95\% highest posterior density (HPD) interval is given as $[-0.50, 0.23]$. However, the resulting Bayes factor $BF_{01}$ in favour of the null hypothesis $H_0:\delta=0$ is given as $BF_{01}=3.71$, which signals moderate evidence for the null hypothesis according to \cite{VanDoorn2019}.

The FBST was conducted with the same Cauchy prior as reference function, that is $r(\delta)=C(0,\sqrt{2}/2)$. Figure \ref{fig:example1fbst} visualises the $e$-value $\overline{\text{ev}}(H_0)$ against $H_0:\delta=0$ as the blue shaded area under the surprise function. Notice that the surprise function is not identical to the posterior, but now equals the ratio $s(\delta)=p(\delta|x)/r(\delta)$ of the posterior distribution $p(\delta|x)$ and the Cauchy prior $r(\delta)=C(0,\sqrt{2}/2)$. 

The $e$-value is obtained via numerical optimisation and integration as $\overline{\text{ev}}(H_0)=0.57$. This shows that only a little more than half of the posterior distribution's parameter values attain higher surprise function values than $\delta_0=0$, which shows that there is not much evidence against $H_0$. Based on the $e$-value against $H_0$ one would therefore not reject $H_0$. Notice however that in contrast to the Bayes factor, confirmation of $H_0$ is not possible.

The $e$-value in favour of $H_0$ is given as $\text{ev}(H_0)=0.43$, and the corresponding $p$-value $ev_0$ based on $k=3$ ($\Theta$ consists of two means $\mu_1$ and $\mu_2$ which are free to vary and the standard deviation, and $\Theta_{H_0}$ consists only of one difference $\mu_1-\mu_2$ which is fixed to the value zero and the standard deviation), $m_0=1.67$, $M_0=-0.13$ and $d_0=||m_0-M_0||^2=3.26$ is computed as 
\begin{align*}
    ev_0 = \chi_2^2(3.26)=0.6463
\end{align*}
Here, $M_0$ was obtained via numerical optimisation and $m_0$ is simply the posterior density's value at $\delta_0=0$. The resulting $p$-value is not significant when the threshold $ev_0 < 0.05$ is applied, so the conclusion is identical to the continuous interpretation of the $e$-value above and $H_0$ is not rejected. Based on the standardized $e$-value against $H_0$, which is $\text{sev}(H_0)=0.0945$, the null hypothesis would not be rejected, too. A standard two-sample t-test would produce a $p$-value of $0.4542$, also producing a non-significant result.

\begin{figure}
	\begin{subfigure}[c]{0.50\textwidth}
		\includegraphics[width=1\textwidth]{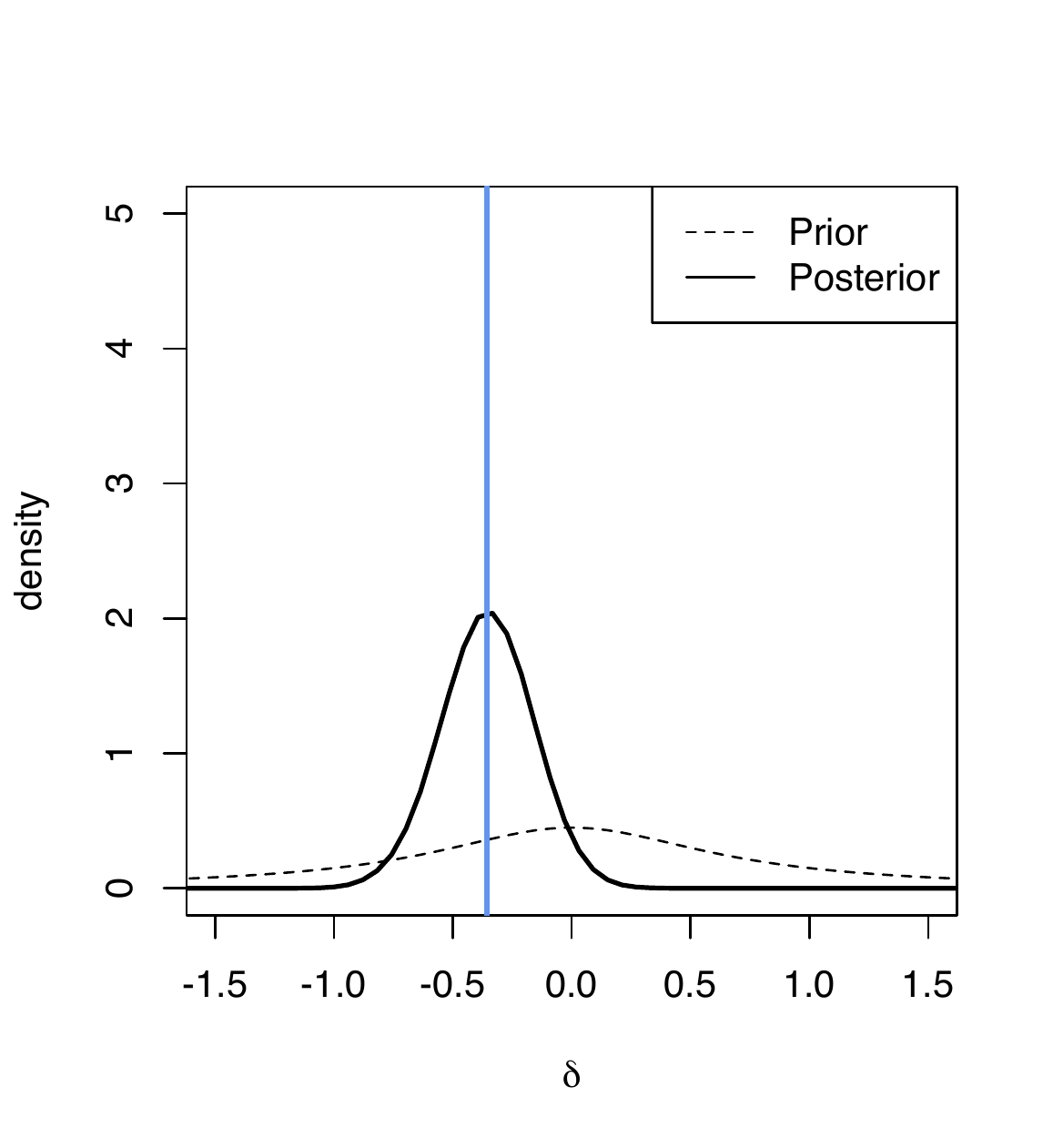}
		\subcaption{Prior-posterior plot of effect size $\delta$ in the Bayesian two-sample t-test for the simulated data set}
		\label{fig:example1bposterior}
	\end{subfigure}
	\begin{subfigure}[c]{0.50\textwidth}
		\includegraphics[width=1\textwidth]{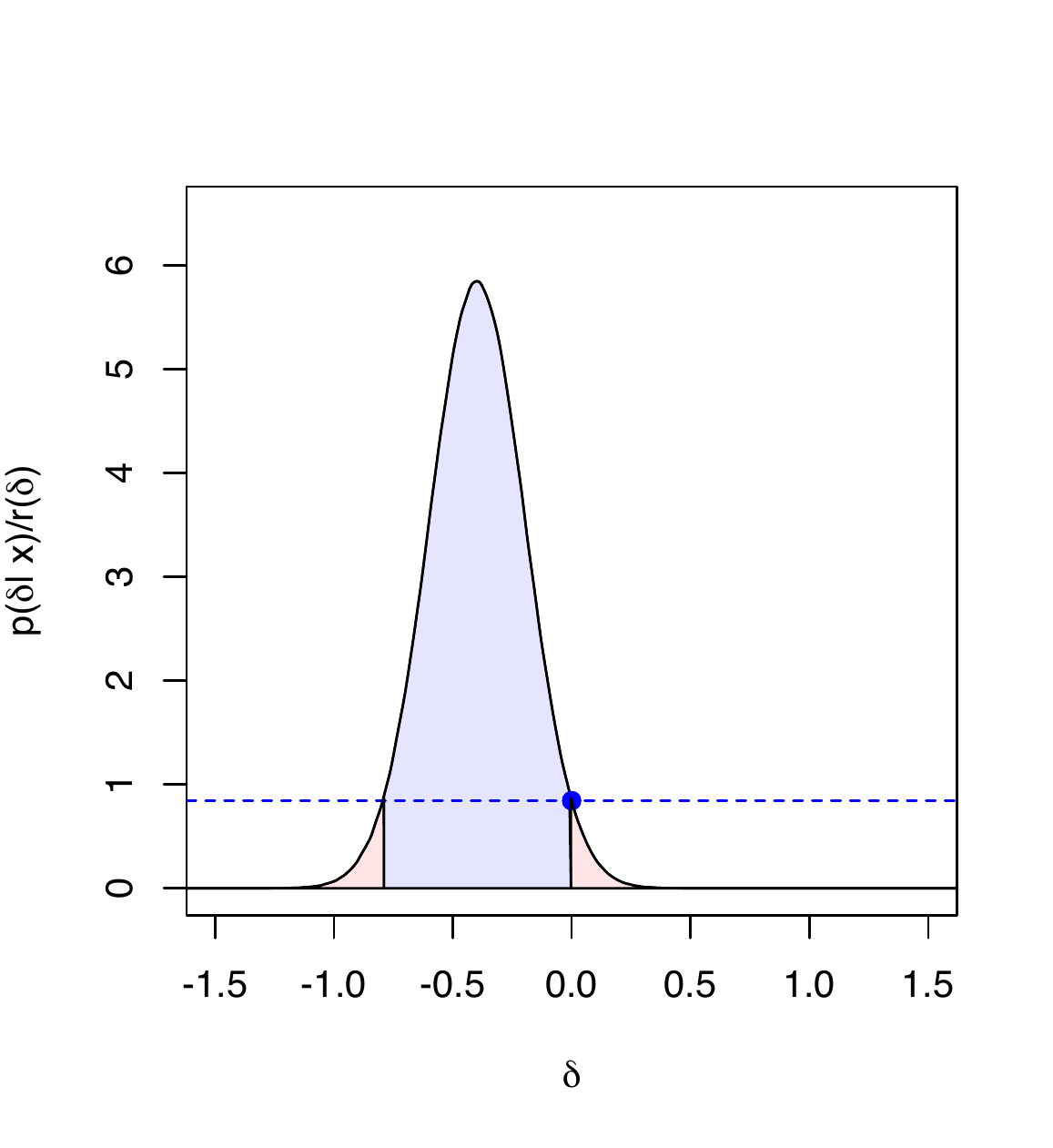}
		\subcaption{FBST for $H_0:\delta=0$ against $H_1:\delta \neq 0$ in the Bayesian two-sample t-test for the simulated data set using $r(\delta)=C(0,\sqrt{2}/2)$}
		\label{fig:example1bfbst}
	\end{subfigure}
	\caption{Prior-posterior plot and FBST for the Bayesian two-sample t-test for the simulated data example}
	\label{fig:example1b}
\end{figure}

Figure \ref{fig:example1b} shows a second Bayesian t-test. This time, data in both groups have been simulated. In the first group, $n=50$ observations were generated according to the $\mathcal{N}(0,1.5)$ distribution and in the second group, $n=50$ observations were generated according to the $\mathcal{N}(0.8,3.2)$ distribution. As a consequence, the resulting true effect size $\delta_t$ according to \cite{cohen_statistical_1988} is given as
\begin{align*}
    \delta_t = \frac{0-0.8}{\sqrt{(1.5^2+3.2^2)/2}} \approx -0.34
\end{align*}
which equals a small effect. The posterior is shown in figure \ref{fig:example1bposterior}.

The resulting Bayes factor $BF_{01}=0.84$ is indecisive ($BF_{10}=1.19$) and the 95\% HPD is given as $[-0.74, 0.02]$. Again, a medium Cauchy prior was used on $\delta$ as previously. Here, the Bayes factor fails to detect the existing small effect.

The FBST was conducted again with the medium Cauchy reference function $r(\delta)=C(0,\sqrt{2}/2)$ and the resulting $e$-value against $H_0$ is given as $\overline{\text{ev}}(H_0)=0.96$ which signals strong evidence against $H_0:\delta=0$. Figure \ref{fig:example1bfbst} visualises the $e$-value again as the blue shaded region under the posterior. 

Considering the $e$-value $\text{ev}(H_0)$ in support of $H_0$, it is obtained as $\text{ev}(H_0)=0.04$, and the corresponding $p$-value $ev_0$ based on $k=2$, $m_0=0.38$, $M_0=-0.35$ and $d_0=0.02$ is calculated as $ev_0=0.01$. Based on a threshold $ev_0 y0.05$ one would, as a consequence, reject the null hypothesis $H_0$ based on the $p$-value $ev_0$ associated with the Bayesian $e$-value $\text{ev}(H_0)$ in support of $H_0$.

\subsection{The FBST in the setting of Bayesian linear regression}
As a second example, we showcase the application of the FBST in the standard Bayesian linear regression model \citep{VanErp2019}. In the example, we use student performance data which were collected by Paulo Cortez and Alice Silva at the University of Minho in Portugal \citep{Cortez2008}. The data can be openly accessed from the University of California-Irvine's machine learning repository \citep{UCIMachineLearningRepository} at \url{http://archive.ics.uci.edu/ml/datasets/Student+Performance}. We use the math performance data which contains math exam scores and multiple predictors from $n=395$ Portuguese students.\footnote{A list of all $30$ predictors available for predicting students' math exam score is given at \url{http://archive.ics.uci.edu/ml/datasets/Student+Performance}, which includes variables like the student's family size, her free time after school or her health status.} Our goal here is to study the influence of the predictors on the first-trimester math grade of each student, which ranges from $0$ to $20$. For illustration purposes, we study the influence of a small subset of the predictors which consists of the gender, the age (ranging from 15 to 22), the time needed to travel to school, the weekly study time and whether the student is in a relationship. The daily travel and weekly study time are measured in four levels where for the daily travel time, 1 = less than 15 minutes, 2 = 15 to 30 minutes, 3 = 30 minutes to one hour, and 4 = more than one hour. For the weekly study time, 1 = less than two hours, 2 = two to five hours, 3 = five to ten hours, and 4 = more than ten hours.

\begin{figure*}[!h]
\centering
\includegraphics[width=1\textwidth]{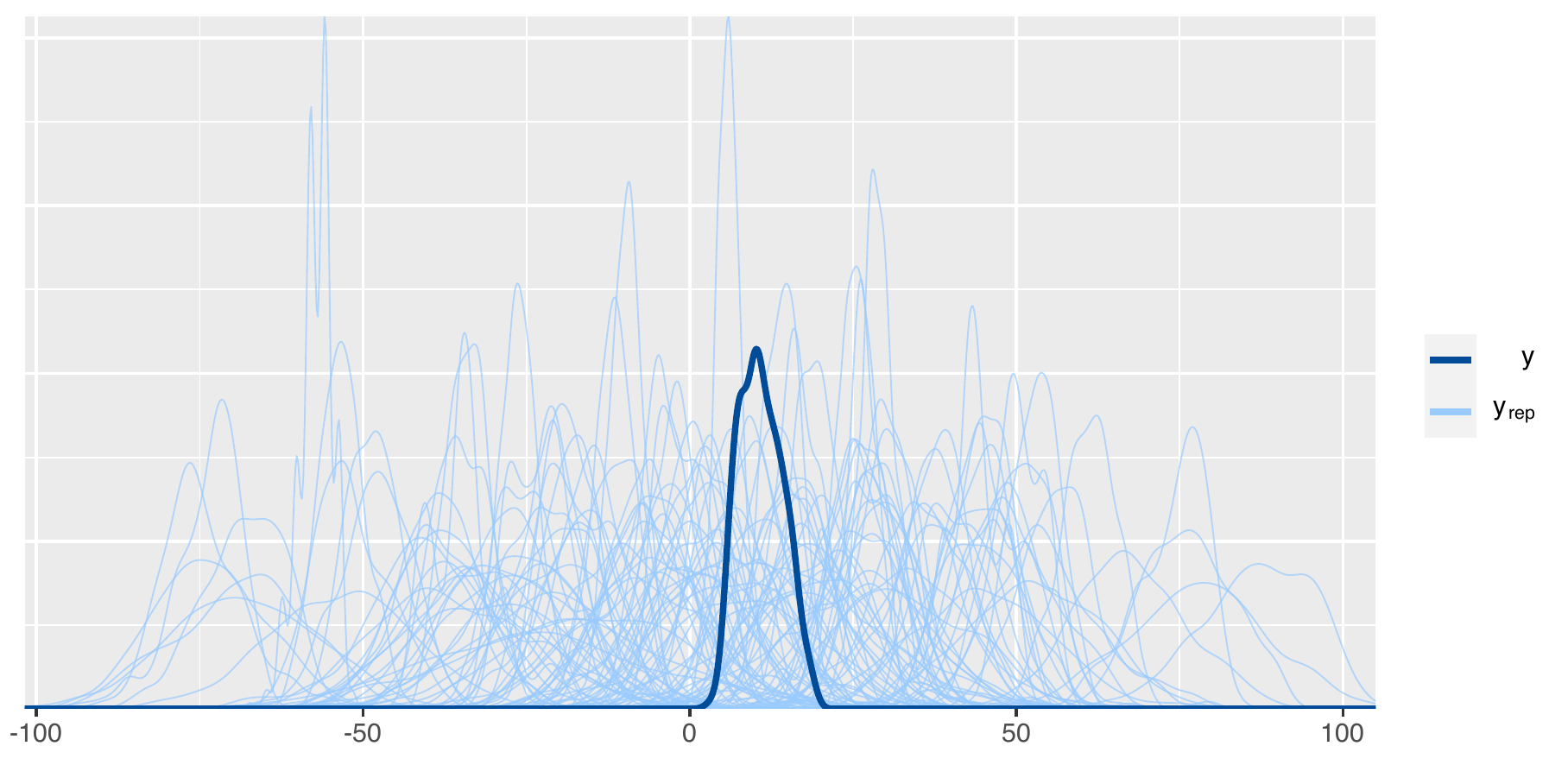}
\caption{Prior predictive distribution for the Bayesian linear regression model of the Student's math performance dataset of Cortez \& Silva (2008)}
\label{fig:example3priorpredictive}
\end{figure*}

For all regression coefficients $\beta$, we choose a normal prior: $\beta_j \sim \mathcal{N}(0,1)$ for $j=1,...,5$. For the intercept, we select $\beta_0 \sim \mathcal{N}(0,10)$, which is the default weakly informative prior, compare \cite{Gabry2020RstanarmPriorsVignette}. For the standard deviation $\sigma$, we choose the default weakly informative $\sigma \sim \exp(1)$ prior, compare also \cite{Gabry2020RstanarmPriorsVignette}. The hyperparameters were selected based on a prior-predictive simulation which is shown in figure \ref{fig:example3priorpredictive}.

Based on the prior predictive distribution, the selected priors are reasonable in the sense that it is plausible that they could have produced the observed data.

\begin{figure*}[!h]
\centering
\includegraphics[width=\textwidth]{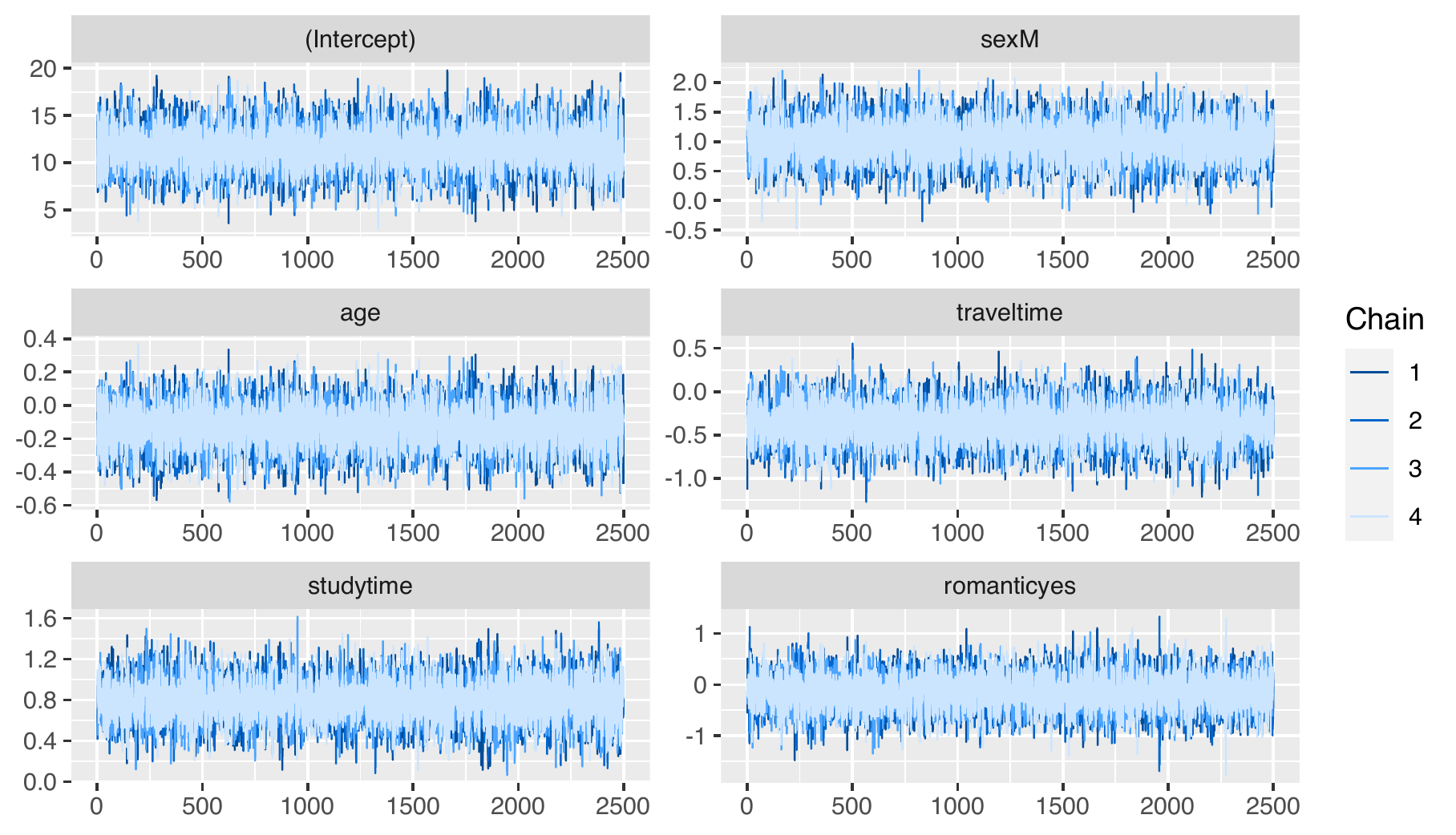}
\caption{Traceplots of the Markov chains for the marginal posteriors of the regression coefficients in the Bayesian linear regression model of the Student's math performance dataset of Cortez \& Silva (2008). \texttt{romanticyes} are the values for the regression coefficients for the predictor relationship when set to one (student is in a romantic relationship), and \texttt{sexM} are the values for the regression coefficients for the predictor sex when set to one (student is male)}
\label{fig:example3traceplots}
\end{figure*}

\begin{center}
\begin{table}
    \centering
  \caption{Posterior estimates of the Bayesian linear regression model of the Student's math performance dataset of Cortez \& Silva (2008)}
  \label{tab:example2}
  \begin{tabular}{llllllr}         
  \toprule
  Estimates & mean  & sd  & 10\% &  50\% & 90\% & $\hat{R}$\\
  \midrule
  (Intercept) & 11.6 & 2.3 & 8.7 & 11.6 & 14.5 & 1.00 \\
  sex       & 1.0 & 0.3 & 0.6 & 1.0 & 1.5 & 1.00\\
  age      & -0.1 & 0.1 & -0.3 & -0.1 & 0.0 & 1.00\\
  traveltime & -0.4 & 0.2 & -0.7 & -0.4 & -0.1 & 1.00\\
  studytime  & 0.8 & 0.2 & 0.5 & 0.8 & 1.1 & 1.00\\
  relationship & -0.2 & 0.4 & -0.6 & -0.2 & 0.3 & 1.00\\
  sigma     & 3.2 & 0.1 & 3.1 & 3.2 & 3.4 & 1.00\\ \bottomrule
  \end{tabular}
\end{table}
\end{center}
Table \ref{tab:example2} shows the resulting posterior estimates based on 2500 posterior parameter draws obtained via the \texttt{rstanarm} R package \citep{Goodrich2020}. Figure \ref{fig:example3traceplots} shows the traceplots of the posterior Markov chain draws for each of the marginal posteriors of the regression coefficients $\beta_j$, $j=0,...,5$. Based on figure \ref{fig:example3traceplots}, all chains are well-behaved and have converged to the posterior. The $\hat{R}$ Gelman-Rubin-shrink factor given in table \ref{tab:example2} is also one for all predictors and signals convergence to the posterior \citep{Gelman1992}.

\begin{figure*}[!h]
\centering
\includegraphics[width=1\textwidth]{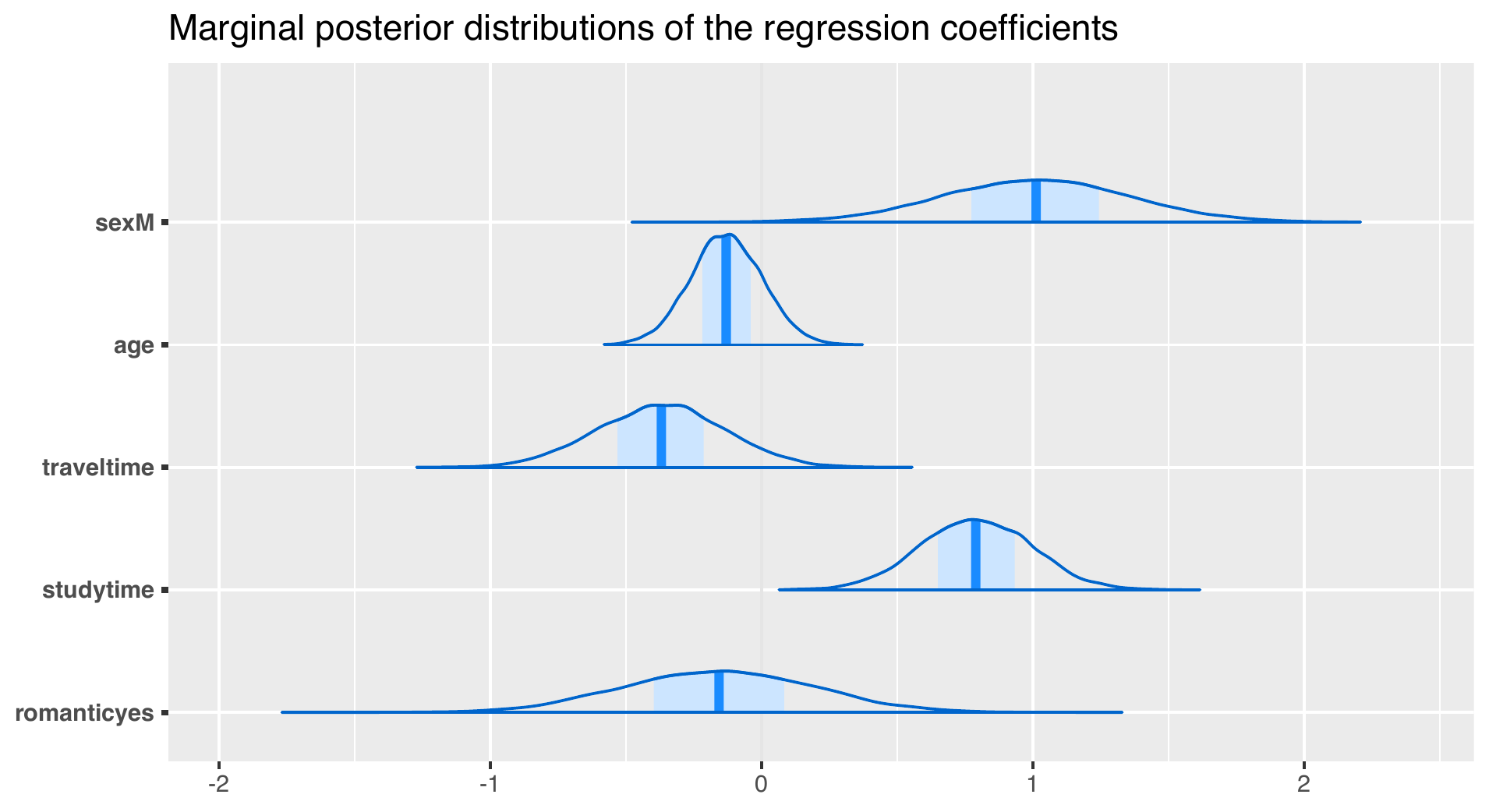}
\caption{Marginal posterior distribution's for the regression coefficients $\beta_1,...,\beta_5$ for the Bayesian linear regression model of the Student's math performance dataset of Cortez \& Silva (2008)}
\label{fig:example3margPost}
\end{figure*}

To apply the FBST, the marginal posterior distributions of each of the regression coefficients $\beta_j$, $j=1,...,6$ are used which are shown in figure \ref{fig:example3margPost}. Based on these posteriors one immediately sees that some covariates influence the first-trimester math grade stronger than others: For example, the coefficient for gender is centred around the value $\beta_j=1$, showing a beneficial influence, while the predictors for relationship and traveltime are shifted towards negative values, indicating that student's in a relationship or with a longer travel time to school perform worse. However, to quantify the evidence against $H_0:\beta_j = 0$ for $j=1,...,5$ we apply the FBST in each case, using a flat reference function $r(\beta)=1$ (so that the surprise function becomes the posterior). For the five predictors, the resulting $e$-values $\overline{\text{ev}}(H_0)$ against $H_0:\beta_j = 0$ and the standardized $e$-values $\text{sev}(H_0)$ are given in table \ref{tab:eValuesPValues}. Based on this continuous quantification of the evidence against $H_0:\beta_j=0$ one would reject the null hypothesis of no influence for the predictors sex and study time. Also, there is some evidence that the travel time plays a role while the age and relationship status are less relevant for predicting first-trimester math performance. The associated $p$-values $ev_0$ with the Bayesian $e$-values $\text{ev}(H_0)$ in support of $H_0:\beta_j=0$ are also shown in table \ref{tab:eValuesPValues}. When the significance threshold $ev_0 < 0.05$ is used, the predictors sex, travel time, study time and relationship are significant.

Notice that the continuous quantification via $\overline{\text{ev}}(H_0)$ is better interpretable: For example, the difference in $p$-values $ev_0 \approx 0.001$ associated with the predictor studytime and $ev_0 \approx 0.002$ associated with the predictor traveltime seems tiny (the same holds for the standardized $e$-values $\text{sev}(H_0)$), but the difference between $\overline{\text{ev}}(H_0)= 1.000$ and $\overline{\text{ev}}(H_0)=0.882$, the Bayesian $e$-values against $H_0$ for the predictors studytime and traveltime, reveals that there is a non-negligible difference between both posterior distributions.
\begin{center}
\begin{table}
    \centering
  \caption{$e$-values $\overline{\text{ev}}(H_0)$, associated $p$-values $ev_0$, and standardized $e$-values $\text{sev}(H_0)$ for the FBSTs of the regression coefficients in the Bayesian linear regression model of the Student's math performance dataset of Cortez \& Silva (2008). Eight-digit precision is used to highlight the differences between the values.}
  \label{tab:eValuesPValues}
  \begin{tabular}{llll}         
  \toprule
  Predictor & $\overline{\text{ev}}(H_0)$ & $ev_0$ & $\text{sev}(H_0)$\\
  \midrule
  sex (male) & 0.996 & 0.00529900 & 0.00000408\\
  age       & 0.658 & 0.22538084  & 0.00495210\\
  traveltime      & 0.881 & 0.00228437 & 0.00069724\\
  studytime & 1.000 & 0.00105736 & 0.00000000\\
  relationship (yes) & 0.346 & 0.01191338 & 0.02464463\\
  \bottomrule
  \end{tabular}
\end{table}
\end{center}

This is also shown in figure \ref{fig:example3margPost}: The marginal posterior distribution of the predictor studytime shows that the regression coefficient of this predictor is certainly positive a posteriori, while the marginal posterior of the predictor traveltime shows a considerable amount of probabiliy mass which indicates that the regression coefficient could also be zero or even positive. Notice also that the $p$-value $ev_0$ (and $\text{sev}(H_0)$) relies on asymptotic arguments which can be questioned if sample sizes are only moderate.

\section{Discussion}
Hypothesis testing remains a substantial statistical method in psychological research. While the problems of null hypothesis significance testing (NHST) are still being debated widely, few appealing alternatives exist to the current status quo. 

In this paper, we provided a tutorial on the Full Bayesian Significance Test (FBST) and the $e$-value, the Bayesian evidence value which was originally developed by \cite{Pereira1999} to test a sharp null hypothesis against its alternative. The foundations and the mathematical theory of the FBST were outlined and it was shown that the FBST is a fully Bayesian alternative to traditional significance tests which rely on $p$-values. There are multiple appealing properties which make the FBST an attractive alternative to NHST and $p$-values: First, the FBST is a formal Bayes rule for an appropriate loss function. Second, it offers a seamless transition from NHST and $p$-values to Bayesian data analysis, which eases the interpretation for practitioners who are accustomed to $p$-values and requires little methodological changes. Third, the FBST is an advanced methodological procedure which can be applied to several areas due to its simple computational basis. Fourth, the FBST can be used with the asymptotic arguments detailed in this paper to use the Bayesian evidence value in support of a sharp null hypothesis to calculate a traditional $p$-value for rejecting the null hypothesis, if desired.

%However, note that this associated $p$-value has a strictly Bayesian interpretation and is not in conflict with the likelihood principle. 

In this tutorial, we showed via two examples of widely used statistical methods in psychological research how the FBST can be used in practice. The first example detailed how to apply the FBST in the setting of the two-sample t-test for a flat and Cauchy reference function. The second example highlighted the FBST in the setting of the Bayesian linear regression model and showed how to test the regression coefficients for the existence of an effect. Notice that both examples were based on completely different statistical models and different software implementations, which shows how easy it is to apply the FBST.\footnote{While the posterior distribution of the Bayesian two-sample t-test of \cite{Rouder2009} was obtained via the \texttt{BayesFactor} package \citep{BayesFactorPackage}, the \texttt{rstanarm} package \citep{Goodrich2020} was used to obtain the posterior for the Bayesian linear regression model.}

However, there are also some limitations of the method: First, analytical solutions are not available in most cases and as a consequence, a minimal requirement to apply the FBST is that the posterior distribution can be obtained via simulation, for example via MCMC sampling. Then, the surprise function can be approximated via kernel estimators or spline-based approaches. Luckily, this is no severe limitation as most realistic Bayesian psychological models are obtained via advanced MCMC sampling techniques in everyday practice \citep{Wagenmakers2010,VanDoorn2019,VanDoorn2020,Kruschke2018}. A more severe limitation of the FBST is that is can not be used to confirm a research hypothesis, in contrast to the Bayes factor or the region of practical equivalence \citep{Kelter2020BayesianPosteriorIndices}. However, the FBST can be generalized into an extended framework which allows for hypothesis confirmation, and this is an active topic of research \citep{Esteves2019, Kelter2020BayesianPosteriorIndices}.

However, the FBST is an innovative method which has, next to its appealing theoretical properties, clearly demonstrated to perform better than frequentist significance testing \citep{Madruga2003, Stern2003,Pereira2008,Stern2020,Kelter2020BayesianPosteriorIndices}. To our best knowledge, it has not been used so far in the psychological sciences and should be of wide interest to a broad range of researchers in psychology. We hope that this paper fosters discussion about the use and suitability of the FBST for psychological research and practice, and enables researchers to apply the FBST to their own data sets and models of interest.

%ignites critical reflection about the discussion about statistical significance, which often is reduced to opposing the Bayes factor to traditional $p$-values.

\medskip

\bibliographystyle{mslapa}
\bibliography{library.bib} 

\newpage

\end{document}